\documentclass[twocolumn]{aa} 
\usepackage{graphicx}
\usepackage{txfonts}
\begin{document}
   \title{Lithium enhancement in X-ray binaries due to stellar rotation}

   \author{T.J. Maccarone\inst{1} 
          \and
	  P.G. Jonker\inst{2}
	  \and
	  A.I. Sills\inst{3}
          }

   \offprints{T.J. Maccarone}

   \institute{Astronomical Institute ``Anton Pannekoek,'' University
              of Amsterdam, Kruislaan 403, Amsterdam, The Netherlands,
              1098 SJ\\ \email{tjm@science.uva.nl} \and
              Harvard-Smithsonian Center for Astrophysics, 60 Garden
              Street, Cambridge, MA 02139, USA; Chandra Fellow \and
              Department of Physics and Astronomy, McMaster
              University, Hamilton, Ontario,L8S 4M1, Canada}

   \date{}

   \abstract{We discuss the high lithium abundances in the secondary
stars of X-ray binaries.  We show that no lithium production in these
stars is necessary, and that the abundances can be explained simply
due to the tidally locked rotation of the stars, which lead naturally to
slower lithium destruction rates.  The differences in abundances of
CVs' secondaries from those of LMXBs had previously been put forth as
evidence that the compact object was related to the lithium abundance,
but this scenario also accounts for the lower lithium abundances in
the secondary stars in cataclysmic variable systems (CVs) than in low
mass X-ray binaries (LMXBs), since these stars have typically lived
much longer before becoming tidally locked short period systems.  We
point out that if this scenario is correct, then the globular cluster
X-ray binaries' donor stars should, as a class, show less lithium
enhancement relative to other stars of the same spectral type in the
clusters than the field X-ray binaries' donor stars show.}

   \maketitle
%
\section{Introduction}

Light element abundances (deuterium, helium and lithium) are a key
tracer of the physical conditions in the early universe, as these are
the only elements produced in the Big Bang.  They provide a key
constraint on the baryon-to-photon ratio at the time of the universe's
formation.  As a result, considerable effort has been invested in
determining the rate at which lithium is destroyed, and in identifying
the stars which have the highest lithium abundances.  A key factor for
understanding how to convert the observed lithium abundances observed
in stars today into primordial values is to understand mechanisms for
lithium production, rather than just lithium destruction.  It has been
noted in several cases that the lithium abundances in X-ray binary
donor stars are substantially larger than the solar value (Martin et
al. 1994; Martin et al. 1996 and references within); this effect is
seen in all X-ray binary secondary stars where there are good
constraints on the lithium abundances.  Since compact objects are
natural sites for the acceleration of relativistic particles or flows,
several mechanisms related directly to the compact object nature of
the systems have been suggested -- spallation in the inner accretion
flow (Martin et al. 1995; Yi \& Narayan 1997), spallation due to the
impact on the stellar atmosphere of high energy neutrons produced in
the inner accretion flow (Guessom \& Kazanas 1999), or interactions
between a relativistic jet and the mass donor (Butt, Maccarone \&
Prantzos 2003), or alternatively, lithium production due to some
unspecified process in stellar coronae (Bildsten \& Rutledge 2000).
With the exception of the last two mechanisms, the transport of the
lithium back into the stellar atmosphere is a non-trivial problem.  In
this paper, we will review the evidence that X-ray binaries' donor
stars have higher than normal lithium abundances.  We will then
demonstrate the similarities with other tidally locked binary stars,
and will discuss the possibility that this lithium enhancement may be
a natural consequence of the slower lithium depletion in tidally
locked binary stars, rather than a consequence of any physics related
to the compact objects.  We will explain, in the context of this
picture, why this enhancement is not seen so strongly in cataclysmic
variable stars (CVs).

\section{Summary of past X-ray binary observations and normal star 
observations}

There are, at the present time, five secondary stars in X-ray
binaries with measured lithium abundances.  The measured abundances,
expressed as, $A({\rm Li}) = {\rm log} N({\rm Li})/N(\rm H) + 12.0$,
range from 2.2 to 3.3, where the solar value is approximately 1.0, so
these stars are above the solar value by a factor of about 20-200.
The actual enhancement relative to what is expected based on normal
single star evolution depends heavily on the unknown stellar age and
on the stellar mass (see e.g. the $\approx$2 dex difference between
the cool stars in the 80 Myr old Pleiades and in the 700 Myr old
Hyades compiled by Ryan \& Deliyannis 1995) The lithium abundance
values of X-ray binary donors are all presented in Martin et
al. (1996), with the exception of that for Cen X-4, which has $A({\rm
Li})$=3.3, and is reported separately (Martin et al. 1994).  Martin et
al. (1996) also note that the ionizing radiation from the X-ray source
might lead to an underestimate of the lithium abundance.

The meteoritic lithium abundance in the solar system has a value of
3.31$\pm$0.04 (Anders \& Gravesse 1989).  This is generally accepted
to be the typical initial lithium abundance of Population I stars
(e.g. Ryan \& Deliyannis 1995).  Hence, even the most lithium-rich of
the X-ray binaries, Cen X-4, which has a lithium abundance equal to
this value, does not require any lithium enhancement of the secondary
star, so long as there has not been substantial lithium depletion in
this star over its lifetime, and so long as the underestimate due to
the lack of accounting for ionization effects is not too severe.

A wealth of literature, both observational and theoretical, is devoted
to the effects of rotationally-induced mixing on lithium depletion in
stellar atmospheres.  It has been suggested that rotating stars should
deplete their initial atmospheric lithium abundances more slowly than
do non-rotating stars, and that this slower depletion of lithium
should be an especially strong effect in tidally locked binaries where
internal shears (which mix the lithium into the inner regions of the
star where it is destroyed) will be prevented from developing due to
the very slow angular momentum losses of the stars (Pinsonneault et
al. 1990).  Furthermore, it has been found in open clusters that the
short-period (i.e. with periods less than 6 days) tidally locked
binary systems have substantially higher lithium abundances than
single stars of the same type, or stars of the same type in wide
binaries; the most lithium-rich stars in open clusters are these
short-period tidally locked binaries, and they often show lithium
abundances rather close to the meteoritic value (see e.g. Soderblom et
al. 1990; Ryan \& Deliyannis 1995).  These results stand in contrast
to the claim of Martin et al. (1994) that, while tidal locking of the
LMXBs might play some role in their enhanced lithium abundances, it
was likely to be a rather minor effect.

At the present time, only a small number of open cluster
tidally-locked binaries have measured lithium abundances (9 appear in
the paper of Ryan \& Deliyannis (1995), and there has not been any
systematic study made since then).  Seven of the nine systems have
$A({\rm Li})>1.8$ and four have $A({\rm Li})$ of at least 3.0,
indicating that the distributions of the X-ray binary systems and the
open cluster systems are in relatively good agreement.  However, the
open cluster systems typically have hotter stars, so it is a better
comparison to look at their overabundances relative to the mean trend
with temperature for the clusters in which they are found; most of
these systems are at least 0.6 dex higher in lithium than the trend
curve in lithium versus effective temperature of the open clusters of
which they are members.  The two Pleiades systems show no measureable
lithium enhancement, but this is a fact which is in agreement with
theoretical models of rotationally induced lithium depletion given the
youth of the Pleides cluster.  

The data is not as conclusive for subgiant stars.  For two subgiants
in the Hyades, lithium is detected at temperatures where there are
only upper limits for stars which are not short period tidally locked
binaries and where the trend line extrapolates to be about 100 times
lower than the detections.  On the other hand, the lithium abundances
are lower than $A({\rm Li})=1.8$ in these two stars. Clearly, more
observations of tidally-locked subgiants are needed to determine
whether tidal locking can account for a substantial fraction of the
lithium excess in X-ray binaries with subgiant donors.

A typical factor of $\sim$5 enhancement in lithium abundance is thus
found for the tidally locked binaries in open clusters.  Furthermore,
it should be noted also that the effects of tidal locking for the
donor stars in X-ray binaries should extend much deeper into the
stellar interior than for those systems compiled by Ryan \& Deliyannis
(1995). The main sequence systems studied by Ryan \& Deliyannis (1995)
all have orbital periods greater than 3 days, while all of the X-ray
binaries except V404 Cyg in the sample have periods less than 1 day
and most have periods less than 10 hours.  Pinsonneault et al. (1989)
studied the lithium depletion as a function of the timescale for
angular momentum transport between the surface and the stellar
interior; they found it to have ``a strong influence'' on lithium
depletion, capable of accounting for 2 dex of variation over the range
they studied.  The model they ran for a solar-type star with weak
angular momentum transport retained A(Li) greater than 2.5 for about
300 Myrs, and greater than 2.0 for approximately 5 Gyrs.  The donors
in X-ray binaries should have essentially no angular momentum
transport within the star; it therefore seems quite reasonable that
they could have more lithium enhancement than in the case of the
wider binaries.

\section{Discussion}
\subsection{Past interpretations}
One of the strongest arguments made in the past for the position that
the enhancement of lithium in X-ray binary donor stars must be related
to the compact object, rather than to the stellar rotation, is the
point that the stellar companions in cataclysmic variables show no
evidence for lithium enhancement (Martin et al. 1995; Yi \& Narayan
1997).  This argument is flawed, though, if one assumes, as is
predicted by stellar theory and borne out by stellar observations,
that lithium is constantly being destroyed at a rather slow rate in
short-period tidally locked binary systems, while it is being
destroyed at a much faster rate in stars with low rotation rates.  The
black hole and neutron star binaries are the end products of evolution
of very massive stars with very short lifetimes.  The progenitors of
the white dwarfs found in cataclysmic variable systems, on the other
hand, have considerably longer lifetimes.  Since the orbital periods
of these systems shrink rather late in their evolution, it is likely
for most CVs that while the white dwarf progenitor was on the main
sequence, the binary was not tidally locked, so the lithium depletion
was proceeding in more or less the normal fashion for isolated or wide
binary stars.  In fact, it is then likely that some of the cataclysmic
variable's donor stars will show mildly enhanced lithium
overabundances, but since a rather large amount of lithium depletion
probably occurs in the first 100 Myrs of stellar evolution (see
e.g. Pinsonneault et al. 1990), CVs' secondaries should be
substantially more lithium depleted than neutron star or black hole
binaries.

On the other hand, Bildsten \& Rutledge (2000) drew attention to the
similarity between X-ray binaries' donor stars and the members of
RsCvn systems.  The stars in RsCvn systems are likely to be tidally
locked short period binaries from rather early on in their
evolution. Bildsten \& Rutledge (2000) had been arguing that the X-ray
emission from quiescent low mass X-ray binaries was primarily from the
stellar coronae of the mass donors, rather than from a low luminosity
accretion flow.  In that context, they argued that since both
quiescent LMXBs and RsCvn stars had enhanced lithium abundances and
X-ray emission(as does at least one white dwarf binary system), that
perhaps both were caused by the coronal activity.  On top of the
arguments of Bildsten \& Rutledge (2000) that the lithium enhancement
is unlikely to be caused by processes related directly to accretion
onto compact objects, we note that the secondary with the highest
lithium abundance, by far, is that in Cen X-4, the only neutron star
X-ray binary whose secondary's lithium abundance has been measured.
Many of the mechanisms for enhancing lithium abundance in X-ray
binaries' secondary stars are related to phenomena which are
considerably stronger in black hole systems than in neutron star
systems -- relativistic jets (the mechanism of Butt et al. 2003) from
black holes are considerably more powerful than those from neutron
stars (see e.g. Migliari et al. 2003), and the radiatively inefficient
accretion flows (the mechanism of Yi \& Narayan 1997) seen from black
hole transients are generally hotter than those seen from neutron star
transients.

\subsection{Interpretation in terms of stellar rotation effects}
In fact, though, it is not necessary to {\it produce} lithium in X-ray
binaries.  We note that a large part of the reason why it is thought
that the donor stars in X-ray binaries are enhanced in lithium is that
they are far more lithium abundant than is the Sun.  For most
elements, substantial deviations from solar abundances are indicative
of some unusual evolutionary history.  However, the solar lithium
abundance at birth was $A({\rm Li})$=3.3, and over the 5 Gigayear
lifetime of the Sun, this value has fallen by a factor of about 200.
Therefore, if lithium depletion is halted in the very high rotation
rate donor stars like the secondaries in X-ray binaries, then it would
not be surprising to see these systems with lithium abundances closer
to the standard value at formation for Population I stars, rather than
at the solar or sub-solar values typically seen in evolved Population
I stars.  An exact quantification of this effect is well beyond the
scope of this paper; calculations of lithium depletion in tidally
locked binaries have not yet been done.  The angular momenta of the
secondary stars in most X-ray binaries are an order of magnitude
higher than the fastest rotators studied by Pinsonneault et
al. (1990), and furthermore, the lithium depletion is further stemmed
due to the very slow angular momentum loss in these systems.  Zahn
(1994) attempted to estimate the lithium depletion as a function of
angular momentum for tidally locked binaries, but extrapolating from
the relation presented in that paper gives a lithium abundance well in
excess of the meteoritic value.  It thus seems most likely that the
secondary stars in X-ray binaries undergo some weak lithium depletion
early in their lifetimes, before the supernova explosion creating the
compact object, and then undergo little or no further lithium
depletion, but considerable further theoretical work should be done to
test this hypothesis.

A possible problem with this mechanism for lithium enhancement is the
mass loss from the stellar envelope during the system's accreting
phase.  We note, though, that the stars for which the Li 6708 Angstrom
line can be observed are the cooler stars among the X-ray binary
counterparts.  As a result, these fall into two categories - subgiants
with relatively long periods whose lithium abundance evolution is
poorly understood both theoretically and observationally (a category
including only V404 Cyg, whose orbital period is 6.5 days), and main
sequence stars with quite short periods -- all the other lithium rich
secondaries are in systems with orbital periods shorter than 1 day
(with most of them being shorter than 10 hours).  It has been shown on
both theoretical and observational grounds (Brocksopp et al. 2004;
Meyer-Hofmeister 2004; Portegies Zwart, Dewi \& Maccarone 2004) that
the shortest period X-ray binaries have the lowest outburst amplitudes
and the lowest mean mass accretion rates among X-ray binaries.  The
theoretical calculations of the expected mass accretion rates from
systems in this orbital period range are of order $3\times10^{-10}$
$M_\odot$ yr$^{-1}$ (King, Kolb \& Burderi 1996), so at least 1 Gyr
would be required before these systems underwent substantial lithium
depletion due to stripping of the outer layers of the stars.
Furthermore, while starting from orbital periods shorter than those of
the open cluster systems which have been studied, LMXBs will typically
require several Gigayears of evolution as tight, tidally locked
binaries before actually coming into Roche lobe contact and starting
accretion (e.g. Kalogera \& Webbink 1998), further prolonging their
lifetimes as lithium rich objects.  Many short period X-ray binaries
should thus have pre-accretion lifetimes of at least a substantial
fraction of a Hubble time, allowing most of them to remain as lithium
rich objects today.

\subsubsection{Observational tests}
We can make one rather strong prediction based on the suggestion that
LMXB secondary stars have enhanced lithium because of reduced
depletion, and rather than because of actual lithium production.  The
mass donors in globular cluster X-ray binaries should, as a class,
show less lithium enhancement when compared with other stars of
similar spectral types in the same clusters than the mass donors in
non-cluster X-ray binaries.  The reason for this is that the X-ray
binaries in globular clusters are formed through dynamical captures
(Clark 1975) -- either tidal interactions (Fabian, Rees \& Pringle
1975) or three-body (or perhaps four-body) exchange interactions
(Hills 1976).  As a result, many of these stars will have undergone
normal stellar evolution for most of their lifetimes before they
became part of a close binary.  It should be noted that there is one
millisecond pulsar in the globular cluster NGC 6397 whose binary
companion shows a lithium enhancement (Sabbi et al. 2003), but it does
not seem likely that good measurements of the lithium abundances of a
large sample of globular cluster X-ray binary secondary stars will be
made with current instrumentation; this task requires both a large
aperture (i.e. greater than that of the HST) and very good angular
resolution (i.e. better than that available from the ground without
adaptive optics).  Plans exist for allowing adaptive optics
measurements in the red part of the optical spectrum in the near
future, so there is some hope for executing this test of rotational
lithium enhancement in X-ray binary secondaries in the near future.

Similarly, the stars in low mass X-ray binaries with rotation powered
pulsars (i.e. high magnetic field neutron stars) as their compact
objects should also be a good test bed for theories of lithium
production; that these systems contain slowly spinning, high magnetic
field neutron stars is probably indicative of a much shorter history
as an accretor; furthermore, these systems are the one class of X-ray
binaries that have never been seen to show relativistic jets (Fender
2005), and the highest energies in their X-ray spectra are much lower
than those seen in the low/hard states of low magnetic field neutron
stars or black holes.

If it can be proven that the lithium abundances in X-ray binaries'
stellar counterparts are the result of preservation of lithium due to
stellar rotation early in the stars' lifetimes, rather than some means
of strong lithium production, this would provide us with a natural
means for determining which neutron stars in binaries were formed in
supernovae, and which were formed in accretion induced collapses.  It
has been suggested in some cases that accretion onto a white dwarf, in
certain mass accretion rate regimes, can lead to the accumulation of
enough mass to allow the white dwarf to exceed the Chandrasekhar
limit, and to become a neutron star (e.g. Bailyn \& Grindlay 1990; van
den Heuvel \& Bitzaraki 1995).  Since these systems will have
undergone a temporary phase as an accreting white dwarf, and hence
they will have had mass donors which were not rapid rotators until
relatively late in their lifetimes, they should have lithium
abundances in their secondary stars which are similar to those of the
cataclysmic variables (i.e. not significantly enhanced relative to
field stars), rather than to those of the other low mass X-ray
binaries.  At the present time, it is rather difficult to test this
hypothesis; the systems suggested to have evolved into neutron stars
through accretion induced collapse are either globular cluster systems
(Bailyn \& Grindlay 1990) or they are systems in which the donor stars
themselves are white dwarfs (van den Heuvel \& Bitzaraki 1995).  In
the former case, the neutron star may have changed partners over its
evolution, while in the latter cases, the expected lithium abundance
is not clear.  Furthermore, at the present time, the measurements of
lithium abundances in the neutron star LMXBs' secondaries are quite
poor; there exists only the detection of lithium in Cen X-4 and an
upper limit in Aql X-1 which is not sufficiently low as to be
constraining (Garcia et al. 1999).

Another apparently natural test would seem to be a measurement of the
$^6{\rm Li}/^7{\rm Li}$ ratio, but the lines are very close together,
and the $^6{\rm Li}$ abundance measurements are typically considerably
lower than the $^7{\rm Li}$ abundances.  To make useful measurements
would require a very long integration on a very large telescope, but
would also require that the system to be observed is nearly face-on,
so that the orbital and rotational broadening of the lines is
minimized.  A more feasible test would be to measure the Be
abundances.  If the lithium overabundances in X-ray binary secondaries
are due to tidal locking preserving lithium, rather than lithium
production, then one would expect the Be abundances to track the
lithium abundances in the same way as they do in other tidally locked
binaries.  Because the strongest Be lines are at 3130 Angstroms, this
would require the source to be nearly directly overhead in order to
minimize the atmospheric absorption.  Fortunately, this is possible
with large planned or existing flexibly scheduled telescopes for two
of the systems with strong lithium enhancements -- V404 Cyg, which is
sometimes nearly directly overhead from the Hobby-Eberly Telescope,
and Cen X-4, which is sometimes nearly directly overhead from the site
of the SALT telescope.  Boron abundance measurements would be
similarly useful, but can be made only from space.

\section{Conclusions}
It is shown here that the tidal locking of the mass donors in X-ray
binaries should result in an enhancement in lithium abundance of a
factor of about 5 when compared with what would be expected from
single stars of the same age and mass.  Tidal locking is therefore
likely to be an important factor in producing the lithium
overabundances seen in X-ray binaries' counterparts.  It could be the
sole factor in the event that the effects of tidal locking are more
important in the tighter X-ray binaries than they are in the tidally
locked binaries which have been observed in open clusters, or if the
X-ray binaries are typically younger than a few Gyrs.

\label{lastpage}
\end{document}